\documentclass[%
reprint,
 amsmath,amssymb,
 aps,
prl,
]{revtex4-1}

\usepackage{graphicx}
\usepackage{dcolumn}
\usepackage{bm}

\begin{document}

\title{Controllable directive radiation of a magnetic dipole above planar metal surface}

\author{Zheng Xi}
\author{Yonghua Lu}
\email{yhlu@ustc.edu.cn}
\author{Peijun Yao}
\author{Wenhai Yu}
\author{Pei Wang}
\author{Hai Ming}
\affiliation{Department of Optics and Optical Engineering,
University of Science and Technology of China,
Heifei 230026, China}

\date{\today}

\begin{abstract}

We report unidirectional radiation of a magnetic dipole above planar metal surface, the radiation direction can be manipulated via changing the distance between the dipole and the surface. This phenomenon is unique for the combination of magnetic dipole and metal surface and does not happen for linear polarized dipole on metal surface or magnetic dipole on dielectric surface. The underlining physics is analytically disclosed by the interference of two orthogonally-oriented dipole component with $\pi/2$ phase lag. A substantially different mechanism of introducing the vectorial nature of the dipole itself to control light emission distinguishes the present scheme from nanoantenna and provides a new degree of freedom in light emission engineering.

\end{abstract}

\pacs{Valid PACS appear here}

\maketitle


The study of dipole emission near planar surface is a fundamental issue in many researches ranging from cavity quantum electrodynamics \cite{Barnes1998}, single molecule fluorescence \cite{Gersen2000}, chemistry \cite{Enderlein1999} and basic antenna theory \cite{Sommerfeld1909}. In the field of nanophotonics, because the electric dipole transition is often dominant over magnetic dipole transition, most of the previous studies have focused on the properties of electric dipole over planar surface \cite{Lukosz1981,Novotny1997,Luan2006,Luan2008,Penninck2010,Inam2011,Kidwai2011} with only a few exceptions \cite{Lukosz1977,Lukosz1977a,Lukosz1979,Karaveli2011}.

Recently, the magnetic counterpart, although assigned different names such as circular polarized dipole\cite{Rodrguez-Fortuo2013} or rotating dipole\cite{Mueller2013}, began to gain much attention \cite{Lee2012,Mueller2013,Rodrguez-Fortuo2013,Seung-YeolLee2013,R1}. It is theoretically conceived and experimentally demonstrated that, when placed above planar surface, the magnetic dipole exhibits attractive property of unidirectional launching of guided modes including surface plasmon polariton modes \cite{Lee2012,Mueller2013,Rodrguez-Fortuo2013}. In these studies, little attention has been paid on the emission properties of the magnetic dipole itself, for example, the control over the dipole emission directivity, which is of crucial importance in the field of nanophotonics. Strategies using optical nanoantenna have been adopted to obtain good emission directivity of the electric dipole \cite{Curto2010,Taminiau2008,KosakoTerukazu2010} while the magnetic dipole is less investigated.

In this Letter, with well-known properties of electric dipole, we study the emission properties of a generalized dipole above various planar surfaces, especially the case for a magnetic dipole above the metal surface both numerically and analytically. Our results reveal that, for a magnetic dipole above the metal surface, the far field radiation pattern is strongly dependent on the dipole-surface distance. Specifically, we show how interference between two dipole components of the magnetic dipole leads to unidirectional emission. Furthermore, for the simple geometry that is so fundamentally considered, the radiation direction sweeps over a wide range in the upper space by slightly changing dipole-metal surface distance without using complex nanoantenna structure. This behavior can be well predicted analytically.
\begin{figure}[b]
\includegraphics[width=3in]{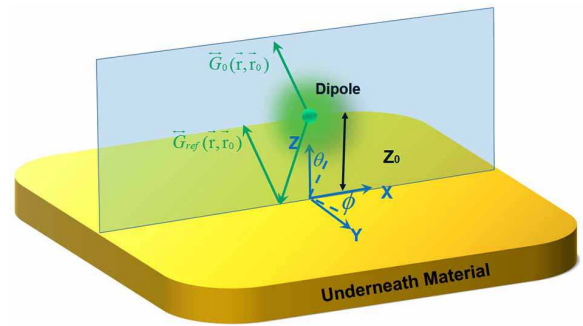}
\caption{(Color Online) Schematic illustration of the considered geometry. A dipole (green dot) is placed at $z_{0}$ above an underneath material. \label{fig:Graph1}}
\end{figure}

Fig.1 presents the schematic geometry of our problem: a dipole oriented within XZ plane is placed at distance $z_{0}$ above a planar metal surface. Generally, the dipole is expressed by a dipole moment $\vec{p}=[p_{x},0,p_{z}]=[1,0,\alpha e^{i\delta}]$. Borrowing the terminology from polarization optics, when $\delta=0$, the expression represents a linearly polarized dipole oriented in XZ plane. When $\alpha=1, \delta=\pi/2$, it is a circular polarized dipole. However, simple calculation reveals it is the same as a magnetic dipole with $\vec{m}=\hat{y}$ \cite{Jackson1998}. For cases other than above, it is an elliptically polarized dipole. This prompts us that a general dipole $\vec{p}$ can be viewed as superposition of two orthogonally oriented linear polarized dipoles with different phase lag, for example, $p_{x}$ and $p_{z}$ here.

The dyadic Green's function method \cite{Jackson1998}, which is a widely used technique in solving problems involving dipole emission near planar surfaces, is adopted to numerically investigate the emission properties. The electric field in the upper and lower space can be written as the summation from two contributions: 1) the free space Green's tensor $\tensor{G}_{0}$  and 2) the reflected (upper space) $\tensor{G}_{ref}$ or transmitted (lower space) Green's tensor $\tensor{G}_{trans}$  or  by the surface \cite{Novotny2006}:
\begin{equation}
\vec{E}{{(\vec{r})}_{up,low}}={{\omega }^{2}}\mu {{\mu }_{0}}({{\tensor{G}}_{0}}(\vec{r},{{\vec{r}}_{0}})+{{\tensor{G}}_{ref,trans}}(\vec{r},{{\vec{r}}_{0}}))\vec{p}
\label{eq:1}
\end{equation}
The reflected or transmitted Green's tensor can be calculated using Fresnel equation by dividing the dipole field into s and p component.The Green's tensor is calculated numerically using OJF Martin's technique to deal with stratified media \cite{Paulus2000}.

In the following, we focus on the emission property in the dipole plane, that is XZ plane in Fig.1 and consider three different cases: a) a $45^{o}$ linear polarized dipole with $\alpha=1,\delta=0$  above a gold surface. b) a magnetic dipole with $\alpha=1,\delta=\pi/2$ above a gold surface and c) a magnetic dipole with $\alpha=1,\delta=\pi/2$ above a dielectric material surface. The dipole is assumed to emit at 500nm and the refractive index of gold is taken from Palik with $n=0.855+1.8955i$ \cite{Palik1998}. The near field electric field $|E|$ plots as well as the far field power pattern with varying dipole-surface distance $z_{0}$ are calculated using dyadic Green's Function method shown in Fig.2.
\begin{figure}
\includegraphics[width=3in]{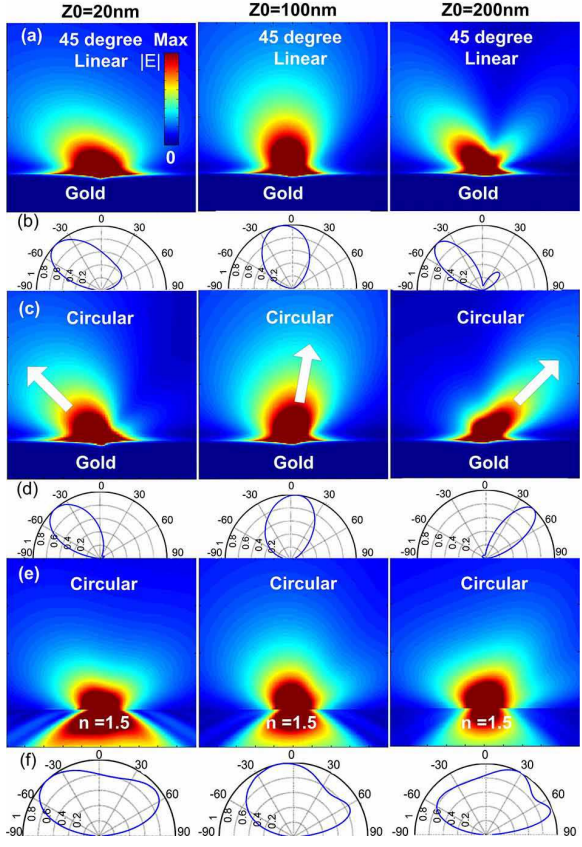}
\caption{(Color Online) Electric field $|E|$ distribution(a,c,e) and Far field power pattern in the XZ plane(b,d,f) of (a,b) $45^{o}$ linear polarized dipole (c,d) magnetic dipole above a gold surface and (e,f) magnetic dipole above $n=1.5$ dielectric material with different dipole surface distances $z_{0}$. The emission wavelength of the dipole is taken to be 500nm and the permittivity of gold is taken from Palik. \label{fig:Graph2}}
\end{figure}

Although the dipole-surface distance $z_{0}$ of each column is identical, the far-field emission patterns differ significantly, especially for the magnetic dipole above the gold surface. Qualitatively, the radiation pattern depends on three important parameters: the phase difference $\delta$  in different dipole component, the reflection coefficient $r_{p}$ from the surface and the dipole-surface distance $z_{0}$. Together, they define the phase acquired by the reflected field. For the $45^{o}$ linear polarized dipole above gold surface, more energy is radiated in the $-\theta$  plane resulting asymmetry in the radiation pattern. The asymmetry itself does not change with different $z_{0}$. Interesting enough, when replaced by a magnetic dipole, the radiation becomes nearly unidirectional and its direction sweeps continuously from $-43^{o}$  to $45^{o}$  with changing dipole-surface distance $z_{0}$. It is also interesting to note the emission pattern narrows meanwhile. This is in sharp contrast to the radiation of a magnetic dipole in free space, which is uniformly distributed in the dipole plane. However, if such a magnetic dipole is placed above a dielectric surface instead of a metal one, the radiation pattern does not change much\cite{Sup}.

To understand and predict the above phenomena, we derive analytical expression for the far field radiation in the upper space. Using angular spectrum representation, the electric field in the XZ upper space far away from the generalized dipole emitter can be written as \cite{Novotny2006}:
\begin{equation}
{{\vec{E}}_{far,up}}(\vec{r})=\frac{{{k}^{2}}}{4\pi {{\varepsilon }_{0}}{{\varepsilon }_{1}}}\frac{e^{i\vec{k}\cdot \vec{r}}}{r}({{p}_{x}}cos\theta {{\Phi }_{2}}-{{p}_{z}}\sin \theta {{\Phi }_{1}}){{\vec{e}}_{\theta }}
\label{eq:2}
\end{equation}
which can be understood clearly as linear superposition of horizontally and vertically oriented dipoles. The $\Phi$ term incorporates the effect of the infinitely large plane below, with $\Phi_{1,2}=e^{-ik{{z}_{0}}cos\theta}\pm {{r}_{p}}e^{ik{{z}_{0}}cos\theta}$ , meaning the total field can be written as the summation of the dipole field itself and its own image weighted by the Fresnel reflection coefficient ${{r}_{p}}=({{n}_{2}}{{k}_{1z}}-{{n}_{1}}{{k}_{2z}})/({{n}_{2}}{{k}_{1z}}+{{n}_{1}}{{k}_{2z}})$  in which  ${{n}_{1}},{{n}_{2}}$ are the refractive index of air and underneath material and ${{k}_{1z}},{{k}_{2z}}$  are component of the wave vector perpendicular to the surface. When the original field and image field are in phase, radiative decay of the dipole is enhanced, when out of phase, the radiative decay is suppressed. The far field radiated power  ${{P}_{far}}(\theta )$ is the proportional to the square of the electric field:
\begin{eqnarray}
{{P}_{far}}(\theta )\propto {{\left| {{p}_{x}} \right|}^{2}}{{\cos }^{2}}\theta {{\left| {{\Phi }_{2}} \right|}^{2}}+{{\left| {{p}_{z}} \right|}^{2}}{{\sin }^{2}}\theta {{\left| {{\Phi }_{1}} \right|}^{2}} \nonumber
\\-\sin \theta \cos \theta ({{p}_{x}}p_{z}^{*}\Phi _{1}^{*}{{\Phi }_{2}}+p_{x}^{*}{{p}_{z}}{{\Phi }_{1}}\Phi _{2}^{*})
\label{eq:3}
\end{eqnarray}
Eq.3 fully captures the angular distribution of the dipole emission power. Besides the horizontal and vertical dipole terms described above, the power radiated by a generalized dipole is also determined by a third interference term. A quick glance at this term, expressed by complex dipole moment, would reveal different behavior for different phase lags introduced by the electric dipole and the magnetic dipole. As we will see in the following, this term is of crucial importance for deciding the radiation pattern.

Let us begin with the case for $45^{o}$ linear polarization $(\alpha=1,\delta=0)$, with dipole components $p_{x}=1,p_{z}=1$. Substitute this into Eq. (3) results in the expression for the far field power pattern:
\begin{eqnarray}
{{P}_{45,far}}(\theta )\propto {{\cos }^{2}}\theta {{\left| {{\Phi }_{2}} \right|}^{2}}+{{\sin }^{2}}\theta {{\left| {{\Phi }_{1}} \right|}^{2}} \nonumber
\\-2\sin \theta \cos \theta (1-{{\left| {{r}_{p}} \right|}^{2}})
\label{eq:4}
\end{eqnarray}
It can be seen in general $P_{45,far}(\theta)\neq P_{45,far}(-\theta)$  resulting asymmetric distribution of the far field radiation pattern. To be specific, we consider contribution of each term in Eq.4. As is mentioned above, the first two terms represent horizontally and vertically oriented dipole field. The electric field itself, however, is of different symmetry for those two dipoles: that is symmetric for horizontal and anti-symmetric for vertical one. This difference is not clear for incoherent dipoles because the power pattern simply adds up. However, when the two dipoles oscillate coherently as in this case, the difference in symmetry of the electric field can introduce additional interference.
As shown in Eq.4, the third interference term is the multiplication of the two dipole electric fields. It is anti-symmetric about $\theta$ (Fig. 3(d)) which is in sharp contrast to the first two terms which are symmetric about $\theta$ (Fig. 3(b),(c)). While the total radiation power is the summation of the three terms, the difference in symmetry indicated by the interference term can introduce asymmetry. As a result, the dipole tends to radiate more in the $-\theta$ plane for all dipole-surface distances (Fig. 3(a)).
\begin{figure}
\includegraphics[width=3in]{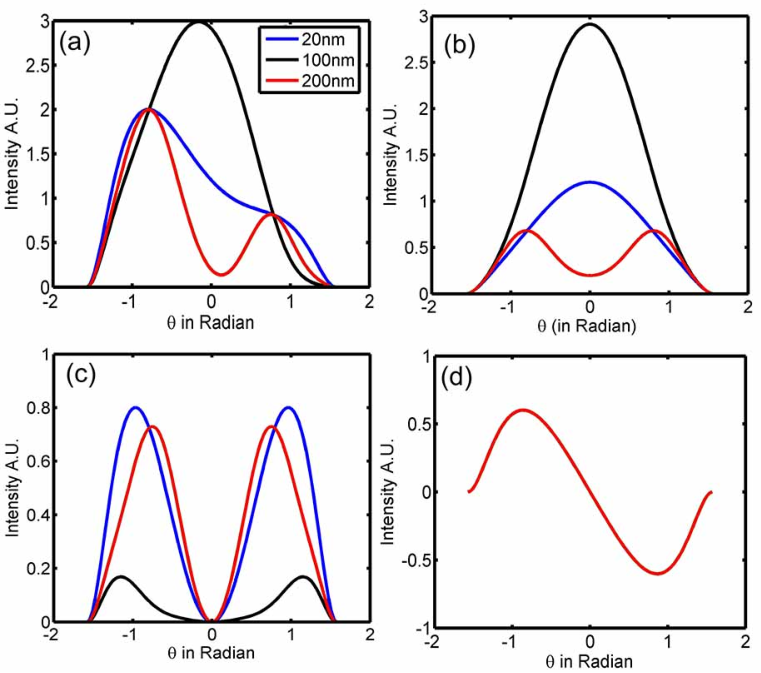}
\caption{(Color Online) Contributions of different terms for a $45^{o}$ linear polarized dipole. (a) The total radiated power in the far field. (b) The cosine term represents a horizontally oriented dipole. (c) The sine term represents a vertically oriented dipole. (d) The interference term between different dipoles. The blue, black and green lines correspond to dipole-surface distance of 20nm, 100nm and 200nm respectively. \label{fig:Graph3}}
\end{figure}

Things become quite different when considering a magnetic dipole $(\alpha=1,\delta=\pi/2)$, with $p_{x}=1,p_{z}=i$ above a gold surface. In this case the far field radiated power pattern can be calculated as:
\begin{eqnarray}
P{(\theta )_{cir,far}} \propto {\cos ^2}\theta {\left| {{\Phi _2}} \right|^2} + {\sin ^2}\theta {\left| {{\Phi _1}} \right|^2} \nonumber
\\- 4\sin \theta \cos \theta \operatorname{Im} ({r_p}{e^{i2k{z_0}\cos \theta }})
\label{eq:5}
\end{eqnarray}
while the first two dipole terms remain unchanged. Of great interest is the third interference term, instead of depending solely on the Fresnel coefficient${{\left| {{r}_{p}} \right|}^{2}}$as in the case for a $45^{o}$ linearly polarized dipole, the interference term now becomes a function of the dipole-surface distance $z_{0}$ while its anti-symmetry property remains unchanged. In Fig. 4(b-d), we plot the contribution of the three term with changing dipole-surface distances $z_{0}$ in Eq. 5. Comparing with the $45^{o}$ linear polarized dipole, a striking difference for the interference term is observed in Fig. 4(d). For the $45^{o}$ linearly polarized dipole, the interference term does not depend on the distance $z_{0}$, therefore the three lines overlap in Fig. 3(d). But for the magnetic dipole, this term varies much when the distance $z_{0}$ changes from 20nm to 200nm. In particular, the third term remains plus in the  $-\theta$ plane for $z_{0}=20nm$ and reverses its sign for $z_{0}=200nm$ (Fig. 4(d)), meaning the radiation is enhanced in one half $\theta$ plane while is suppressed in the other resulting unidirectional emission. Moreover, the summation of this term causes different power distribution in $\theta$ plane for different $z_{0}$. This means that by controlling the dipole-surface distance $z_{0}$, the emission pattern of a magnetic dipole can be swept over a wide angle. Naturally, because of the symmetry consideration, when the magnetic dipole moment flips its sign, or the chirality changes in the language of a circular polarized dipole, the sweeping direction changes accordingly. This is very different from the antenna based light emission control in which case specific antenna resonance is involved to carefully tune the required phase\cite{Curto2010,Taminiau2008,KosakoTerukazu2010}. In this situation, it is the interference between the dipole itself results in directional light emission and dynamic tuning.
\begin{figure}
\includegraphics[width=3in]{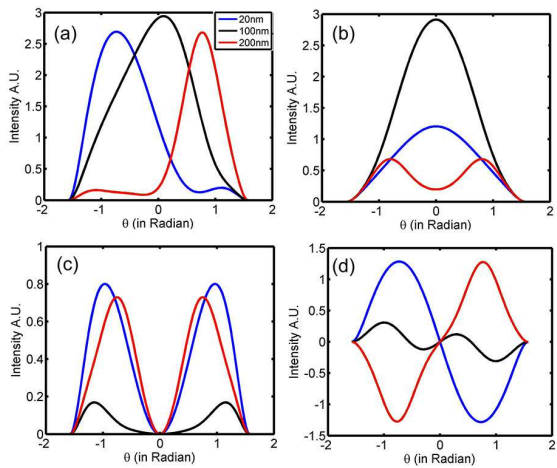}
\caption{(Color Online) Contributions of different terms for a magnetic dipole. (a) The total radiated power in the far field. (b) The cosine term represents a horizontally oriented dipole. (c) The sine term represents a vertically oriented dipole. (d) The interference term between different dipoles. The blue, black and green lines correspond to dipole-surface distance of 20nm, 100nm and 200nm respectively. \label{fig:Graph4}}
\end{figure}

When the magnetic dipole is placed above a normal dielectric material with refractive index n=1.5 instead of gold, the expression for the far field power is the same with Eq. 5, except that the reflection Fresnel coefficient $r_{p}$ is real and can be exacted out from the imaginary part as:
\begin{eqnarray}
P{{(\theta )}_{cir,far,die}}\propto {{\cos }^{2}}\theta {{\left| {{\Phi }_{2}} \right|}^{2}}+{{\sin }^{2}}\theta {{\left| {{\Phi }_{1}} \right|}^{2}} \nonumber
\\-4\sin \theta \cos \theta {{r}_{p}}\sin (2k{{z}_{0}}\cos \theta )
\label{eq:6}
\end{eqnarray}
To effectively affect the radiation pattern, the magnitude of the interference term should be large enough to be comparable to the first two terms. However, as can be seen from Eq.5 and Eq.6, the magnitude of the third term is determined by the magnitude of $r_{p}$. For gold surface considered above, the reflection is about 10 times larger than the dielectric at around $\theta=\pm45^{o}$ . For small $r_{p}$ as in the case for dielectric substrate, the addition of the first two terms approximately equals 1 while the inclusion of the third term adds little modulation, therefore the radiation pattern does not change much.

Finally, we would like to point out the radiation pattern can also be controlled using different values of $\alpha$ and $\delta$  while keeping the dipole-surface distance $z_{0}$ constant to get the required phase. The magnetic dipole in reality can be implemented by considering Zeeman effect for small molecules \cite{AtkinsPW1997} or even molecule with circular polarized luminescence \cite{Petoud2007}. Besides, intra-4f electronic transistions of lanthanide ions such as trubalent erbium which is commonly used from fluorescent lighting and telecommunication amplifiers shows strong optical-frequency magnetic dipole moment \cite{Karaveli2011}.

In conclusion, using rigorous Green's tensor method, we have shown directional emission tuning of a magnetic dipole using a simple metal surface. By slightly changing the dipole-surface distance, the radiation maximum can be swept over a wide range in the upper space. The sweeping direction changes for dipole with different chirality.

These results highlight a simple scheme to engineer the magnetic dipole emission near metal surface, which opens up new design for magnetic-light interactions. Moreover, the inclusion of refractive index changing material such as nonlinear or electro-optical crystals between the dipole -metal surface which is very sensitive to light intensity may provide active control over the light emission.

This work is supported by National Key Basic Research Program of China (2012CB921900 and 2012CB922003), Key Program of National Natural Science Foundation of China (61036005) and National Natural Science Foundation of China (11274293, 61177053).

\end{document}